\documentclass[pra,
twocolumn]{revtex4}
\usepackage{optidef}

\usepackage{graphicx,bm,natbib,upgreek,amsmath,mathrsfs,accents}
\usepackage{amsbsy}
\usepackage{amsfonts}

\usepackage[colorlinks = true,
            linkcolor = blue,
            urlcolor  = blue,
            citecolor = blue,
            anchorcolor = blue]{hyperref}

\usepackage{amsthm}

\usepackage[dvipsnames]{xcolor}

\usepackage{graphicx}
\usepackage{subfigure}

\def\<{\langle}
\def\>{\rangle}
\newcommand{\mytick}[2]{\includegraphics[width={#2}pt]{{#1}}}
\def\tabletick{\vspace{0pt}\mytick{tick}{9}}
\def\tablecross{\vspace{0pt}\mytick{cross}{9}}
\def\yes{\tabletick}
\def\no{\tablecross}

\begin{document}

\title{Quantum key distribution with non-ideal heterodyne detection: \\
composable security of discrete-modulation continuous-variable protocols}
\author{Cosmo Lupo}
\affiliation{Dipartimento di Fisica, Politecnico di Bari, 70126 Bari, Italy\footnote{Previous affiliation: Department of Physics and Astronomy, University of Sheffield, S37RH Sheffield, UK}}

\author{Yingkai Ouyang}
\affiliation{Department of Electrical and Computer Engineering, National University of Singapore, Singapore}

\begin{abstract}
Continuous-variable quantum key distribution exploits coherent measurements of the electromagnetic field, i.e., homodyne or heterodyne detection.
The most advanced security proofs developed so far relied on idealised mathematical models for such measurements, which assume that the measurement outcomes are continuous and unbounded variables.
As physical measurement devices have finite range and precision, these mathematical models only serve as an approximation.
It is expected that, under suitable conditions, the predictions obtained using these simplified models are in good agreement with the actual experimental implementations. 
However, a quantitative analysis of the error introduced by this approximation, and of its impact on composable security, have been lacking so far.
Here we present a theory to rigorously account for the experimental limitations of realistic heterodyne detection.
We focus on collective attacks, and present security proofs for the asymptotic and finite-size regimes, the latter within the framework of composable security.
In doing this, we establish for the first time the composable security of discrete-modulation continuous-variable quantum key distribution in the finite-size regime.
Tight bounds on the key rates are obtained through semi-definite programming and do not rely on a truncation of the Hilbert space.
\end{abstract}

\maketitle

\section{Introduction}
Quantum key distribution (QKD) is the art of exploiting quantum optics to distribute a secret key between distant authenticated users.
Such a secret key can then be used as a one-time pad to achieve unconditionally secure communication.
First introduced in the $80$'s by Bennett and Brassard \cite{BB84}, QKD is now at the forefront of quantum science and technology. 
By encoding information into the quantum electromagnetic field, QKD enables provably secure communication through an insecure communication channel, a task known to be impossible in classical physics.
This contrasts with standard and post-quantum cryptography, which are based on computational assumptions and do not guarantee long-term security. In fact, future advancements in theoretical computer science or computational power (including quantum computing) may jeopardize the security of these schemes.

To travel the route from fundamental physics to future technologies, we need to account for the trade-off between the rate of key generation of the protocol, its security, and the feasibility and robustness to experimental imperfection.
The highest standards of security and robustness are those of device-independent QKD, but are achieved at the cost of a reduced key rate. 
Here we focus on continuous-variable (CV) QKD, within the device-dependent approach, which allows for feasible implementations with much higher key rates. Our goal is to improve the robustness of CV QKD to experimental imperfections and practical limitations.
For a recent review of device-independent QKD and CV QKD we refer to Ref.~\cite{Review2020}.

CV QKD denotes a family of protocols where information is carried by the phase and quadrature of the quantum electromagnetic field. 
A variety of protocols exist that differ in how the quadratures encode this information \cite{GG02,Weedbrook,Lev11,Furrer,Koashi}.
However, when it comes to decoding, all CV QKD protocols exploit coherent measurements of the field, i.e., either homodyne or heterodyne detection \cite{Ferraro}.
The strategic importance of CV QKD indeed relies on this choice of measurement, as homodyne and heterodyne detection are mature, scalable, and noise-resilient technologies. 
This is in contrast with discrete-variable architectures, that require bulky, high-efficiency, and low-noise single-photon detectors \cite{Lo}.

When modeling a CV QKD protocol, it is customary to describe its measurement outcomes as continuous and unbounded variables. %
In these models, homodyne detection measures one quadrature of the field, and heterodyne detection provides a joint measurement of both quadrature and phase \cite{Ferraro}.
These simplified models are powerful mathematical tools due to their continuous symmetry.
Two fundamental theoretical results rely on this symmetry: the \textit{optimality of Gaussian attacks} \cite{Wolf,Navascues,Garcia} and the \textit{Gaussian de Finetti reduction} \cite{Lev2017}. 
However, this symmetry is not exact and is broken by real-world physical devices. In fact, in actual experimental implementations, homodyne and heterodyne detection yield digital outcomes and have a finite range \cite{Jouguet,mioQKD}.
While it is expected that, in some limit, the idealised measurement models describe actual physical devices well, up to now a quantitative analysis of this approximation was lacking.
In particular, it was not known how to quantify the impact of these non-idealities on the secret key rate.

In this work we finally fill this important conceptual gap and present a theory to quantify the secret key rates obtained in actual QKD protocols that exploit actual measurement devices.
Up to now, only a handful of results were available in this direction. Furrer \textit{et al.}~considered digitalised homodyne for a protocol based on distribution of entangled states \cite{Furrer}, and Matsuura \textit{et al.}~considered a binary encoding using coherent states, homodyne detection, and a test phase exploiting heterodyne \cite{Koashi}.
However, in both cases the key rates do not converge to the asymptotic bounds obtained in Refs.~\cite{Lev2017,Lut19,Lev19,Lev21}, which are believed to be optimal for ideal detection. 
In contrast, our results converge to these optimal bounds when the non-idealities are sufficiently small.

We focus on discrete-modulation (DM) protocols, where the sender prepares coherent states whose amplitudes are sampled from a discrete ensemble.
We establish the security against collective attacks in both the asymptotic and non-asymptotic regime, the latter within the framework of composable security \cite{Canetti}. 
This contrasts with previous works on DM CV QKD \cite{Lut19,Lev19,Lut21,Lev21}, which only considered the asymptotic limit of infinite channel uses.
Our composable security proof allows us to quantify the security of QKD in the practical scenario where the number of signal exchanges is finite, and QKD is used a sub-routine of an overarching cryptography protocol.
Although collective attacks are not the most general attacks, they are known to be optimal, up to some finite-size corrections, through de~Finetti reduction \cite{Lev2017,posts,RennerCirac}
%
While we focus on heterodyne detection, the same approach may be as well applied, with some modifications, to homodyne detection.

\section{Structure of the paper and summary of results}

We introduce DM CV QKD with non-ideal heterodyne detection in Section \ref{Sec:model} and review its asymptotic security in Section \ref{Sec:asymptotic}. 
We discuss using a data-driven approach to approximate infinite-dimensional states with ones with finite-dimensional support in Section \ref{Sec:cutoff},
and in Section \ref{Sec:cont} calculate corresponding corrections to our secret key rate by using a continuity argument.

We bound the secret key rates in three different settings with increasing complexity, where in each setting we find the optimal values using linear semi-definite programming.
In the first setting (Section \ref{Sec:SDP}), the semi-definite programs are still over infinite dimensional quantum states, and knowledge of their optimal values would allow one to determine the secret key rate in the asymptotic limit.
In the second setting (Section \ref{Sec:finitedim}), we map the infinite dimensional semi-definite programs of Section \ref{Sec:SDP} into finite-dimensional ones, the latter of which can be solved numerically without truncating the Hilbert space.
This gives us a way to exactly numerically evaluate the secret key rate in the asymptotic limit.
In the third setting (Section \ref{Sec:finsize}), within a composable security framework, we generalize the theory of asymptotic QKD to non-asymptotic QKD; we show how perturbations to the semi-definite programs in Section \ref{Sec:finitedim} depend on the number of channel uses, and prove that these perturbations vanish when the number of channel uses becomes arbitrarily large.
This result allows us to estimate the secret key rate of a non-asymptotic DM CV QKD scheme with composable security.

Explicit examples are discussed in Section \ref{Sec:QPSK}, for the case of Quadrature Phase Shift Keying (QPSK).
These examples suggest that, in the limit of vanishing non-idealities in heterodyne measurement and growing number of channel uses, the secret key rate of DM CV QKD approaches the highest rate possible.
Conclusions and potential future developments are discussed in Section \ref{Sec:end}.

Table \ref{tab:comparison} compares our results with previous works that also presented security analysis of CV QKD protocols.
We only consider works that obtained a tight estimation of the key rate.
The encoding of classical information in quantum signals may happen through either a continuous modulation (CM) or a discrete modulation (DM). In this work we consider DM, which reflects what is actually done in experiments.
We obtain our security proof within the framework of composable security, which is the gold standard in cryptography; composable security permits a quantitative assessment of the security of QKD, including when the QKD protocol is a subroutine of an overarching communication protocol.
We consider a realistic model of actual heterodyne detection, instead of the ideal model used in previous works.
Our numerical calculation of the lower bound on the secret key rate is exact, as we do not need to impose an arbitrary cutoff of the Hilbert space.

\begin{table}[t!]
    \centering
    \begin{tabular}{c||c|c|c|c}
                             & Encoding & Composable & Heterodyne & Key rate \\
         \hline
         Ref.~\cite{Lev2017} & CM         & \yes        & Ideal      & Exact   \\
         Ref.~\cite{Lev19}   & DM         & \no         & Ideal      & Approx.    \\
         Ref.~\cite{Lut19}   & DM         & \no         & Ideal      & Approx. \\
         Ref.~\cite{Lut21}   & DM         & \no         & Ideal      & Exact   \\
         Ref.~\cite{Lev21}   & DM         & \no         & Ideal      & Exact   \\
         This work           & DM         & \yes        & Realistic  & Exact 
    \end{tabular}
    \caption{Comparison of our results with previous security analyses of CV QKD. 
    We only include works that provided a tight estimate of the key rates.
    Encoding: the protocol considered has continuous (CM) or discrete (DM) modulation. 
    Composable: the security analysis is in the framework of composable security.
    Heterodyne: the security analysis assumes an ideal or realistic model for heterodyne detection.
    Key rate: the estimation of the key rate is exact or obtained through numerical approximation.}
    \label{tab:comparison}
\end{table}

\section{The model}\label{Sec:model}
We consider one-way QKD where one user (conventionally called Alice) prepares quantum states and sends them to the other user (called Bob), who measures them by heterodyne detection.
The transmission is through an insecure quantum channel that may be controlled by an adversary (called Eve). 
This general scheme defines a \textit{prepare \& measure} (PM) protocol.
In this work we focus on DM CV QKD protocols where, on each channel use, Alice prepares a coherent state $|\alpha\rangle$ whose amplitude is sampled from an $M$-ary set, $\{ \alpha_x \}_{x=0,\dots,M-1}$ with probabilities $\mathcal{P}_x$. This defines Alice's $M$-ary random variable $X$.
An example is quadrature phase shift keying (QPSK), obtained for $M=4$ and setting $\alpha_x = \alpha i^x$, $\mathcal{P}_x = 1/4$.

In order to prove the security of these protocols, we need to consider a different, though formally equivalent, scenario where a bipartite quantum state $\rho_{AB}$ is distributed to Alice and Bob, of which Eve holds a purification.
This kind of setting defines an \textit{entanglement-based} (EB) protocol.
It is sufficient to prove the security of the EB protocol, from which the security of the PM protocol follows.
In the EB protocol, the state $\rho$ is a two-mode state, where $a$, $a^\dag$ and $b$, $b^\dag$ are the annihilation and creation operators for Alice and Bob, respectively.
The EB representation of DM CV QKD protocols is discussed in detail in Ref.~\cite{Lev21}. 
In this work we focus on {\it collective attacks}, which are identified by the assumption that, over $n$ uses of the quantum channel, the state factorises and
has the form $\rho_{AB}^{\otimes n}$.
In the following, we indicate as $\rho_B = \text{Tr}_A ( \rho_{AB} )$ the reduced state on Bob side. To make the notation lighter, we will sometimes drop the subscripts $AB$ or $B$ when the the meaning is clear from the context.

On the receiver's side, Bob measures by applying heterodyne detection.
Ideally, heterodyne detection is a joint measurement of the field's quadrature ($q$) and phase ($p$), whose output can be described as a complex variable $\beta = (q + i p)/\sqrt{2}$.
Ideal heterodyne detection, applied on a state $\rho$, would yield a continuous and unbounded output, with probability density $\frac{1}{\pi} \, \langle \beta | \rho |\beta \rangle$, where $|\beta\rangle$ is the coherent state of amplitude $\beta$.
In contrast, actual experimental realisations of heterodyne detection have measurement outcomes that are confined to a finite region in phase space, $\beta \in \mathcal{R}(R)$, and hence have finite range. 
Here we assume that the region $\mathcal{R}(R)$ is defined by the condition $q, p \in [-R,R]$, for some $R>0$.
Furthermore, the measurement outputs are digital, such that each quadrature takes $d$ values, with each value corresponding to a unique $\log{d}$-bit string. This is obtained by binning the values of $q \in [-R,R]$ into $d$ non-overlapping intervals. For simplicity, we consider intervals of equal size,
\begin{align} \label{intervals}
\mathcal{I}_{j} = [ - R + 2(j-1)R/d , - R + 2jR/d ] \, ,
\end{align}
for $j=1,\dots, d$. 
The output $j$ is then associated to the event $q \in \mathcal{I}_{j}$, which, in turn, we identify by the central value 
\begin{align}
q_j = - R + (j-1)R/d + jR/d \, .
\end{align}
The same digitisation, when applied to both $q$ and $p$, yields a description of actual heterodyne detection as a measurement with $d^2$ possible outputs. 
This defines Bob's variable $Y$, which is a discrete random variable and assumes $d^2$ values. These discrete values can be conveniently labeled using the central points of each interval, i.e., 
\begin{align}
\beta_{jk} = ( q_j + i p_k)/\sqrt{2} \, .
\end{align}

If Bob obtains the average state $\rho_B$, then the probability of measuring $\beta_{jk}$ is
\begin{align}
    P_{jk} = \int_{\beta \in \mathcal{I}_{jk}} 
    \frac{d^2\beta}{\pi} 
    \langle \beta| \rho_B |\beta \rangle \, ,
\end{align}
where the complex interval $\mathcal{I}_{jk}$ is defined in such a way that 
$\beta \in \mathcal{I}_{jk}$ if and only if $q \in \mathcal{I}_{j}$ and $p \in \mathcal{I}_{k}$, and $d^2\beta = \frac{1}{2}\, dq dp$.
Finally, there is a non-zero probability 
\begin{align} \label{prange}
    P_0(R) = 1 - \int_{\beta \in \mathcal{R}(R)} 
    \frac{d^2 \beta}{\pi}  
    \langle \beta| \rho_B |\beta \rangle 
\end{align}
of an inconclusive measurement, when the amplitude lies outside the measurement range.

\section{Asymptotic security of CV QKD}\label{Sec:asymptotic}
In the limit that $n \to \infty$, the secret key rate (i.e., the number of secret bits that can be distilled per transmission of the signal) is given by the Devetak-Winter formula \cite{Devetak}:
\begin{align}
    r_\infty = \xi I(X;Y) - \chi(Y;E)_\rho \, ,
\end{align}
where $I(X;Y)$ is the mutual information between Alice and Bob, and $\chi(Y ; E)_\rho$ is the Holevo information (quantum mutual information) between Bob and Eve (here we assume reverse reconciliation on Bob's data, which is optimal for long-distance communication). 
The factor $\xi \in (0,1)$ accounts for the sub-unit efficiency of error correction.
While $I(X;Y)$ only depends on $X$ and $Y$, 
$\chi(Y;E)_\rho$ also depends on the quantum information held by Eve, which in general cannot be estimated directly.
Fortunately, the property of extremality of Gaussian states \cite{Navascues,Garcia} allows us to write the upper bound
\begin{align}
    \chi(Y;E)_{\rho} \leq f_\chi(\gamma_A(\rho),\gamma_B(\rho),\gamma_{AB}(\rho)) \, ,
\end{align}
where $f_\chi$ is a known function of the covariance matrix (CM) elements (see Appendix \ref{App:Holevo})
\begin{align}
\gamma_A(\rho) & := \frac{1}{2} \mathrm{Tr}[ ( a^\dag a + a a^\dag )  \rho ] \, , \\
\gamma_B(\rho) & := \frac{1}{2} \mathrm{Tr}[ ( b^\dag b + b b^\dag ) \rho ] \, , \\
\gamma_{AB}(\rho) & := \frac{1}{2} \mathrm{Tr}[ ( a^\dag b^\dag + a b ) \rho ] \, .
\end{align}

In conclusion, estimating the CM suffices to obtain a universal upper bound on the Holevo information, which holds for collective attacks in the limit of $n \to \infty$.
The asymptotic key rate is thus bounded as
\begin{align}\label{rate1}
    r_\infty \geq \xi I(X;Y) - f_\chi(\gamma_A(\rho),\gamma_B(\rho),\gamma_{AB}(\rho)) \, .
\end{align}
Since $f_\chi$ is an increasing function of $\gamma_A$ and $\gamma_B$, and a decreasing function of $\gamma_{AB}$ \cite{Lev2015}, estimating upper bounds on $\gamma_A$, $\gamma_B$ and a lower bound on $\gamma_{AB}$ suffices to bound the asymptotic key rate.
In practical realisations of CV QKD, where the parameter $\gamma_A$ is known by definition of the protocol, one only needs to bound $\gamma_B$ and $\gamma_{AB}$.

\section{Photon-number cutoff}\label{Sec:cutoff}
The technical difficulties in the analysis of CV QKD are due to the fact that the quantum information carriers reside in a Hilbert space with infinite dimensions. 
To overcome this issue we need to impose a cutoff in the Hilbert space. 
As we do not want to impose such a cutoff in an arbitrary way, we follow a data-driven approach.
Define the following operators on Bob's side:
\begin{align}
W_R = \int_{|\beta|^2 > R^2} 
\frac{d^2 \beta}{\pi} \, |\beta \rangle \langle \beta| \, ,
\end{align}
and
\begin{align}
V_R = \sum_{n > R^2} |n\rangle \langle n| \, ,
\end{align}
where $|n\rangle$ is the Fock state with $n$ photons.
Renner and Cirac noted that \cite{RennerCirac} 
\begin{align}
V_R \leq 2 W_R \, .
\end{align}

From the experimental data, Bob can estimate the probability $P_0(R)$ as in Eq.~(\ref{prange}).
Note that 
\begin{align}
W_R \leq \int_{|\beta|^2 \not\in \mathcal{R}(R)} 
\frac{d^2 \beta}{\pi} \, |\beta \rangle \langle \beta| \, ,
\end{align}
from which we obtain
\begin{align}\label{chain1}
\mathrm{Tr}( V_R \rho_B ) \leq 2 \mathrm{Tr}( W_R \rho_B )
\leq 2 P_0(R) \, .
\end{align}
This shows that the probability that Bob receives more than $2R^2$ photons is no larger than $2 P_0(R)$.
The gentle measurement lemma \cite{Winter} then yields
\begin{align}
    \| \rho_{AB} - \tau_{AB} \|_1 \leq 2 \sqrt{ 2 P_0(R) }\, ,
\end{align}
where
\begin{align} \label{taudef}
    \tau_{AB} = 
    \frac{ (I \otimes \Pi) \rho_{AB} (I \otimes \Pi) }{\text{Tr}(\Pi \rho_{B}) } \, ,
\end{align}
is a normalised state with finite-dimensional support, and
\begin{align}
\Pi = I - V_R = \sum_{n \leq R^2} |n\rangle \langle n| 
\end{align}
is the projector onto the subspace with up to $N = \lfloor R^2 \rfloor$ photons, and $\| \, \cdot \, \|_1$ is the trace norm.
In conclusion, though $\rho$ is generic, an experimental estimation of the probability $P_0(R)$ allows us to determine the proximity of $\rho$ to a state with finite-dimensional support.

\section{Continuity of the Holevo information}\label{Sec:cont}
In the EB representation, the two-mode state $\rho_{AB}$ is measured, on Bob's side, by heterodyne detection.
In general, $\rho_{AB}$ resides in a Hilbert space with infinite dimensions. However, as discussed above, it is close in trace norm to the state $\tau_{AB}$ in Eq.~(\ref{taudef}).
Note that $\tau_{AB}$ has support in a space with $\smash{M \times \lfloor R^2+1 \rfloor}$ dimensions.

The Holevo information is a continuous functional of the state. By applying the continuity bound of Shirokov we obtain \cite{Shirokov} 
\begin{align}
    \chi(Y;E)_\rho \leq \chi(Y;E)_\tau + \delta \, ,
\end{align}
where (in this paper we put $\log \equiv \log_2$, and $\ln$ denotes the natural logarithm)
\begin{align}
\delta = \delta' \log{d^2} + 2 (1+\delta') \log{(1+\delta')} - 2 \delta' \log{\delta'} \, ,
\end{align}
with $\delta' = \| \rho_{AB} - \tau_{AB} \|_1 \leq 2 \sqrt{ 2 P_0(R) }$. 

This implies that, by paying a small penalty in the key rate, we can replace $\rho$ with the finite-dimensional state $\tau$. 
We thereby obtain the following bound on the asymptotic key rate,
\begin{align} \label{rinfty}
    r_\infty \geq \xi I(X;Y) - f_\chi(\gamma_A(\tau),\gamma_B(\tau),\gamma_{AB}(\tau)) - \delta \, .
\end{align}

By comparing with Eq.~(\ref{rate1}), we note that this bound depends on the CM of $\tau$.
However, $\tau$ is only a mathematical tool and does not describe the state that is prepared and measured in the experimental realisation of the protocol. The only state that is physically accessible is $\rho$.
Below we show how we can estimate the CM of $\tau$ by measuring $\rho$ by heterodyne detection.
In particular, our goal is to find an upper bound on $\gamma_B(\tau)$ and a lower bound on $\gamma_{AB}(\tau)$.

\section{Semi-definite programming}\label{Sec:SDP}
In the EB representation, Alice prepares the two-mode state
\begin{align}
|\Psi \rangle_{AA'} = \sum_{x=0}^{M-1} \sqrt{\mathcal{P}_x} \, |\psi_x\rangle_{A} \otimes |\alpha_x\rangle_{A'} \, .
\end{align}
Alice keeps the mode $A$ and sends $A'$ to Bob.
The vectors $|\psi_x\rangle$ are mutually orthogonal and span an $M$-dimensional subspace of Alice's mode $A$. 
Note that Alice's reduced state is
\begin{align}
    \rho_A 
    = \sum_{x,x'=0}^{M-1} \sqrt{\mathcal{P}_x \mathcal{P}_{x'}} \, \langle \alpha_{x'} | \alpha_x\rangle
    |\psi_x\rangle \langle \psi_{x'}| =: \sigma 
    \, .
\end{align}

The equivalence with the PM protocol is obtained by noticing that a projective measurement of $A'$ in the basis $\{ | \psi_x \rangle \}_{x=0,\dots,M-1}$ prepares the mode $A$ in the coherent state $|\alpha_x\rangle$ with probability $\mathcal{P}_x$.
A good choice for the vectors $|\psi_x\rangle$'s is presented in Ref.~\cite{Lev21}.

Our goal is to bound the key rate using the data collected by Alice and Bob, where Bob's measurement is modeled as realistic heterodyne detection with finite range and precision.
We follow the seminal ideas of Refs.\ \cite{Lut19,Lev19} and achieve this by semi-definite programming (SDP).
As an example, we apply linear SDP, as done in Ref.~\cite{Lev19}, to bound the CM of the state $\tau$, but we remark that our theory can also apply to non-linear SDP as in Ref.~\cite{Lut19}.

Let $\rho_B(x)$ be the state received by Bob given that Alice sent $|\alpha_x\rangle$. Alice and Bob can experimentally estimate the probability mass distribution
\begin{align}
    P_{jk|x}
    & = 
    \int_{\mathcal{I}_{jk}} 
    \frac{d^2\beta}{\pi} \langle \beta | \rho_B(x) | \beta \rangle 
    \, ,
\end{align}
which can be used as a constraint in the SDP that we later formulate. 
We can also consider linear combinations of the parameters $P_{jk|x}$, which obviously are also experimentally accessible. 
Here we consider the quantities
\begin{align}
v & := \sum_{j,k=1}^d |\beta_{jk}|^2 P_{jk}  \, , \\
c & :=  
\sum_{x=0}^{M-1} \mathcal{P}_x \sum_{j,k=1}^d \frac{\bar \alpha_x \beta_{jk} + \alpha_x \bar\beta_{jk} }{2} \, P_{jk|x}  \, ,
\end{align}
(where $\bar{\ }$ denotes complex conjugation) which are the expectation values of the variance and the covariance between Alice's and Bob's variables.

Note that $v = \mathrm{Tr}(\mathcal{V} \rho)$ and $c = \mathrm{Tr}(\mathcal{C} \rho)$ are the expectation values of the operators
\begin{align}
    \mathcal{V} & = 
    \sum_{j,k=1}^d |\beta_{jk}|^2 \int_{\mathcal{I}_{jk}} \frac{d^2\beta}{\pi} \, |\beta \rangle \langle \beta| \, , \\
    \mathcal{C} & = \frac{1}{2}
    \sum_{x=0}^{M-1} \sum_{j,k=1}^d
    \bar \alpha_x \beta_{jk} 
    |\psi_x\rangle \langle \psi_x | 
    \otimes 
    \int_{\mathcal{I}_{jk}} 
    \frac{d^2 \beta}{\pi} \, 
    |\beta \rangle \langle \beta| + \mathrm{h.c.} \, .     \label{qomx34}
\end{align}

Similarly, from Eq.~(\ref{prange}), the quantity $1-P_0(R) = \text{Tr}(\mathcal{U} \rho)$ is the expectation value of the operator
\begin{align}
    \mathcal{U} = \int_{\beta \in \mathcal{R}(R)} \frac{d^2\beta}{\pi} \, |\beta \rangle \langle \beta| 
    \, .
\end{align}

Denote as $\tilde \gamma_{B}(\tau)$ the optimal value of the semi-definite program
\begin{maxi}
{\rho_B \ge 0}{ \frac{1}{2} \mathrm{Tr}[ \Pi ( b^\dag b + b b^\dag ) \Pi \rho_B ] } {}{}
\addConstraint{ \text{Tr} ( \mathcal V \rho_B ) }{\leq v}{}
\addConstraint{ \text{Tr} ( \mathcal{U} \rho_B ) }{\geq 1-P_0(R)}{}
\addConstraint{ \text{Tr} (\rho_B)  }{= 1}{}. \label{SDP-upperbound}
\end{maxi} 
Taking into account normalisation, we obtain the upper bound on $\gamma_{B}(\tau)$,
\begin{align}
    \gamma_{B}(\tau) \leq \frac{ \tilde \gamma_{B}(\tau) }{\text{Tr}(\Pi \rho_B)} \leq \frac{\tilde \gamma_{B}(\tau)}{1-2P_0(R)} \, .
\end{align}

Similarly, consider the optimal value $\tilde\gamma_{AB}(\tau)$ of the semi-definite program
\begin{mini}
{\rho_{AB} \ge 0}{ \frac{1}{2} \mathrm{Tr}[ ( a^\dag \Pi b^\dag \Pi + a \Pi b \Pi ) \rho_{AB} ] }{}{}
\addConstraint{ \text{Tr}[ (I\otimes\mathcal{V})  \rho_{AB} ]  }{\leq v}{}
\addConstraint{ \text{Tr}( \mathcal C \rho_{AB} )  }{\geq c}{}
\addConstraint{ \text{Tr} [(I\otimes\mathcal{U}) \rho_{AB} ] }{\geq 1-P_0(R)}{}
\addConstraint{ \mathrm{Tr}_B( \rho_{AB} )  }{= \sigma}{}
\addConstraint{ \text{Tr}(\rho_{AB})  }{= 1}{}, \label{SDP-lowerbound}
\end{mini} 
from which we obtain the lower bound
\begin{align}
    \gamma_{AB}(\tau) \geq \frac{ \tilde \gamma_{AB}(\tau) }{\text{Tr}[(I\otimes \Pi) \rho_{AB}]} \geq \tilde \gamma_{AB}(\tau) \, .
\end{align}

Note that the projector $\Pi$ appears in the objective functions but not in the constraints. For this reason, we cannot simply replace $\rho$ with $\tau$, and the optimal values of the semi-definite programs remain defined in an infinite dimensional Hilbert space.
However, when numerically solving these semi-definite programs, we find solutions of the form $\Pi \rho_B \Pi$ and $(I \otimes \Pi) \rho_{AB} (I \otimes \Pi)$. This suggests that the presence of the projector operator $\Pi$ in the objective function suffices to make the problem effectively finite-dimensional (see the Appendix \ref{App:SDP} for further detail).
To numerically evaluate the optimal values of these semi-definite programs, we derive the corresponding dual programs, which are more efficient to evaluate, and detail this in Appendix \ref{App:SDP}.

\section{Finite-dimensional SDP} \label{Sec:finitedim}
In this section we obtain from (\ref{SDP-upperbound}) and (\ref{SDP-lowerbound}) two semi-definite programs that are defined in a finite-dimensional Hilbert space.
We do this by replacing the constraints appearing in (\ref{SDP-upperbound}) and (\ref{SDP-lowerbound}) with weaker constraints.
This represents no loss of generality, as our goal is to obtain an upper bound on $\gamma_B(\tau)$ and a lower bound on $\gamma_{AB}(\tau)$. 
We express the new semi-definite programs in terms of the normalised state $\tau_{AB}$, defined in Eq.~(\ref{taudef}), which has support in the finite-dimensional subspace containing no more than $N = \lfloor R^2 \rfloor$ photons.

First consider the semi-definite program in (\ref{SDP-upperbound}). %
Note that, since $\mathcal{V}$ is positive semi-definite, we have 
\begin{align}
\text{Tr}( \Pi \mathcal{V} \Pi \rho_B ) \leq \text{Tr}( \mathcal{V} \rho_B ) \, .    
\end{align}
Therefore, the condition $\text{Tr} ( \mathcal V \rho_B ) \leq v$ implies $\text{Tr} ( \mathcal V \Pi \rho_B \Pi ) \leq v$.
Taking into account the fact that the trace of $\Pi \rho_B \Pi$ is larger than $1-2P_0(R)$ (from Eq.~(\ref{chain1})), we obtain the following constraint: 
\begin{align}
\text{Tr} ( \mathcal V \tau_B )
& = \frac{\text{Tr} ( \mathcal V \Pi \rho_B \Pi)}{\text{Tr} ( \Pi \rho_B \Pi)} \\
& \leq \frac{\text{Tr} ( \mathcal V \Pi \rho_B \Pi)}{1-2P_0(R)} 
\leq \frac{v}{1-2P_0(R)} \, .
\end{align}

Also note that the constraint $\text{Tr} ( \mathcal{U} \rho_B ) \geq 1-P_0(R)$ can be rewritten as $\text{Tr} ( (I-\mathcal{U}) \rho_B ) \leq P_0(R)$. As $I-\mathcal{U}$ is positive semi-definite, this constraint can be replaced with $\text{Tr} ( (I-\mathcal{U}) \Pi \rho_B \Pi ) \leq P_0(R)$.
Applying the same argument as above, we obtain the constraint
\begin{align}
\text{Tr} ( (I-\mathcal{U}) \tau_B  ) \leq \frac{P_0(R)}{1-2P_0(R)} \, ,
\end{align}
which in turn implies
\begin{align}
\text{Tr} ( \mathcal{U} \tau_B  ) \geq 1 - \frac{P_0(R)}{1-2P_0(R)} \, .
\end{align}

Putting all this together, (\ref{SDP-upperbound}) can be replaced with the finite-dimensional semi-definite problem:
\begin{maxi}
{\tau_B \ge 0}{ \frac{1}{2} \mathrm{Tr}[  ( b^\dag b + b b^\dag ) \tau_B] } {}{}
\addConstraint{ \text{Tr} ( \mathcal V \tau_B ) }{\leq \frac{v}{1-2P_0(R)}}{}
\addConstraint{ \text{Tr} ( \mathcal{U} \tau_B) }{\geq 1 - \frac{P_0(R)}{1-2P_0(R)}}{}
\addConstraint{ \text{Tr} ( \tau_B)  }{= 1 }{}. \label{SDP-upperbound-finite}
\end{maxi} 

Consider now (\ref{SDP-lowerbound}). 
Note that the operator $\mathcal{C}$ is bounded,
\begin{align}
    \| \mathcal{C} \|_\infty 
    & =
    \sup_{x} \left\|
    \sum_{j,k=1}^d
    \frac{\bar \alpha_x \beta_{jk} +  \alpha_x \bar \beta_{jk}}{2} 
    \int_{\mathcal{I}_{jk}} 
    \frac{d^2 \beta}{\pi} \, 
    |\beta \rangle \langle \beta| \right\|_\infty \\
    & \leq \sup_{x,j,k} \frac{\left| \bar \alpha_x \beta_{jk} +  \alpha_x \bar \beta_{jk}\right|}{2} 
    \left\|
    \int_{\beta \in \mathcal{R}(R)} 
    \frac{d^2 \beta}{\pi} \, 
    |\beta \rangle \langle \beta| \right\|_\infty \\
    & \leq \frac{1}{2} \sup_{x,j,k} \left| \bar \alpha_x \beta_{jk} +  \alpha_x \bar \beta_{jk}\right| \, ,
\end{align}
where $\| O \|_\infty = \sup_\psi \frac{|\langle \psi | O | \psi \rangle|}{\langle \psi | \psi \rangle}$ denotes the operator norm.

This observation allows us to express the constraint in terms of the state $\tau_{AB}$ instead of $\rho_{AB}$ by introducing a small error,
\begin{align}
    \left| \text{Tr}( \mathcal C \rho_{AB} ) - \text{Tr}( \mathcal C \tau_{AB}) \right| 
    & = \left| \text{Tr}( \mathcal C (\rho_{AB} - \tau_{AB} ) ) \right| \\
    & \leq \| \mathcal{C} \|_\infty \| \rho_{AB} - \tau_{AB} \|_1 \\
    & \leq 2 \sqrt{ 2 P_0(R) } \| \mathcal{C} \|_\infty \, ,
\end{align}
where the first inequality follows from the general property that $| \text{Tr}(O O')| \leq \| O \|_\infty \|O'\|_1$, for any pair of Hermitian operators $O$, $O'$.

In conclusion, we replace (\ref{SDP-lowerbound}) with the finite-dimensional semi-definite problem:
\begin{mini}
{\tau_{AB} \ge 0}{ \frac{1}{2} \mathrm{Tr}[ ( a^\dag b^\dag + a b ) \tau_{AB} ] }{}{}
\addConstraint{ \text{Tr}[ (I\otimes\mathcal{V})  \tau_{AB} ]  }{\leq \frac{v}{1-2P_0(R)} }{}
\addConstraint{ \text{Tr}( \mathcal C \tau_{AB} )  }{\geq c - 2 \sqrt{ 2 P_0(R) } \| \mathcal{C} \|_\infty}{}
\addConstraint{ \text{Tr} [(I\otimes\mathcal{U}) \tau_{AB} ] }{\geq 1 - \frac{P_0(R)}{1-2P_0(R)} }{}
\addConstraint{ \mathrm{Tr}_B( \rho_{AB} )  }{= \sigma}{}
\addConstraint{ \text{Tr}(\tau_{AB})  }{= 1 }{}. \label{SDP-lowerbound-finite}
\end{mini}

\section{Non-asymptotic regime}\label{Sec:finsize}

Entropic uncertainty relations are often used to establish the security of QKD in the non-asymptotic regime \cite{EUR2012}.
In particular, they have been applied successfully in CV QKD by Furrer \textit{et al.}~\cite{Furrer}. 
Unfortunately, this elegant method does not yield a tight bound on the key rate for CV QKD. Quoting Leverrier \cite{Lev2017}:
\begin{quote}
\textit{“This~[CV QKD]~protocol can be analyzed thanks to an entropic uncertainty relation, but~[...]~this approach does not recover the secret key rate corresponding to Gaussian attacks in the asymptotic limit of large $n$, even though these attacks are expected to be optimal.”} 
\end{quote}
In the same paper, Leverrier showed that the Asymptotic Equipartition Property (AEP) \cite{AEP} is better suited for CV QKD as it converges to the secret key rate corresponding to Gaussian attacks in the asymptotic limit.

As we show below, the theory developed in the previous sections can be extended to the non-asymptotic regime where a finite number $n$ of signals is exchanged between Alice and Bob. To achieve this goal, we need to make two main modifications to our theoretical analysis. 

The first modification accounts for the finite-size correction to the entropic functions appearing in the asymptotic rate in Eq.~(\ref{rinfty}). These corrections can be computed using the AEP \cite{AEP}:
\begin{align} 
    r_n & \geq \xi I(X;Y) - f_\chi(\gamma_A(\tau),\gamma_B(\tau),\gamma_{AB}(\tau)) - \delta \nonumber \\
    & \phantom{=}~- \frac{\Delta(d,\epsilon_\text{s})}{\sqrt{n}} 
    + \frac{ 2 \log{( \sqrt{2} \epsilon_\text{h} )} }{n} \, ,
    \label{r_n}
\end{align}
where the additive term $\Delta$ can be bounded as \cite{XXX}
\begin{align} 
\Delta(d,\epsilon_\text{s}) \leq 4(1 + \log{d} ) \sqrt{\log{(2/\epsilon_\text{s}^2)}} \, ,
\end{align}
and $\epsilon_s$ is the entropy smoothing parameter.
Furthermore, Eq.~(\ref{r_n}) also includes a term due to privacy amplification, characterised by the hashing parameter $\epsilon_\text{h}$.
The corresponding key is secure up to probability $\epsilon = \epsilon_\text{s} + \epsilon_\text{h}$
(see Ref.~\cite{AEP} for more details).

Invoking the AEP is not sufficient to analyse the non-asymptotic regime. 
In order to achieve composable security in the non-asymptotic regime, we also need to provide confidence intervals for the channel parameters that are not known exactly but obtained through parameter estimation. 
Our second modification to our theory takes this into account, and we discuss this further below. 
Providing confidence intervals for parameter estimation is a difficult problem in CV QKD because the variables measured in ideal homodyne or heterodyne detection are unbounded.
This problem was solved by Leverrier \cite{Lev2017} by exploiting a continuous symmetry of heterodyne detection for CV QKD protocol with Gaussian-modulation. 
Unfortunately, discrete modulation occurs on a finite range and does not have a continuous symmetry. 
Hence, Leverrier’s approach cannot by applied to any CV protocol with discrete modulation.
In our work, since we consider non-ideal heterodyne detection (which is bounded), we are able to compute confidence intervals for all the relevant parameters of the communication channel.
Therefore, although the AEP can be applied to previous asymptotic security proofs (e.g.~Refs.~\cite{Lev21,Lev19,Lut19,Lut21}), our work is the first one to allow for a composable analysis of parameter estimation for CV QKD protocol with discrete modulation.

\subsection{Parameter estimation: confidence intervals}

The second modification arises because the parameters $v$, $c$, and $P_0(R)$, which enter the semi-definite programs, need to be estimated from experimental data. In the non-asymptotic regime, these estimates are subject to statistical errors due to finite-size fluctuations. To account for this, we need to compute confidence intervals for these quantities for any finite $n$. It is sufficient to consider one-sided confidence intervals, as the parameters enter the semi-definite programs in constraints expressed through inequalities. 
Following the approach of Ref.~\cite{Lev2015}, we assume that parameter estimation is performed after error correction. This allows Alice and Bob to use all their raw keys for both parameter estimation and key extraction.

First consider the variance parameter $v$. 
Given $n$ signal transmissions, Bob obtains from his measurements a string of quadrature and phase values, $q_1^B, q_2^B, \dots, q_n^B$ and $p_1^B, p_2^B, \dots, p_n^B$. His best estimate for $v$ is
\begin{align}
    \hat v = \frac{1}{n} \sum_{i=1}^n \frac{(q_i^B)^2 + (p_i^B)^2}{2} \, .
\end{align}
In the scenario of collective attacks, this is the sum of $n$ i.i.d.\ variables, with each variable taking values in the interval $[0,R^2]$.
We can then obtain a confidence interval for $v$ using the additive Chernoff bound.
For any $\delta_v >0$,
\begin{align}\label{ev}
    \text{Pr} \left\{ \hat v < v - \delta_v \right\} \leq \exp{\left[ - n D\left( \frac{v-\delta_v}{R^2} \left\| \frac{v}{R^2} \right. \right) \right]}  \, ,
\end{align}
where $D(a \| b) = a \ln{\frac{a}{b}} + (1-a) \ln{\frac{1-a}{1-b}}$ is the relative entropy.
Note that, for $p<1/2$, we have 
\begin{align}\label{KLbound}
D(p-\epsilon \| p) > \frac{\epsilon^2}{2p(1-p)} \, ,     
\end{align}
which yields
\begin{align}
    \text{Pr} \left\{ \hat v < v - \delta_v \right\}
    & \leq \exp{\left( - \frac{n \delta_v^2}{2R^2 v (1-v/R^2)} \right)}  \\
    & \leq \exp{\left( - \frac{n \delta_v^2}{2R^2 v} \right)} =: \epsilon_v \, .
    \label{ev1}
\end{align}

To obtain a confidence interval for the covariance parameter $c$ we apply the Hoeffding bound.
Let us denote as $q^A_1, q^A_2, \dots, q^A_n$ and $p^A_1, p^A_2, \dots, p^A_n$ the raw data collected by Alice. 
The best estimate for $c$ is
\begin{align}
    \hat c = \frac{1}{n} \sum_{i=1}^n \frac{q^A_i q^B_i + p^A_i p^B_i}{2} \, .
\end{align}
This quantity is the sum of $n$ i.i.d.\ variables, with each variable chosen from the interval $[-A R , A R]$, where $A = \max_x \{ |\mathsf{Re}(\alpha_x)|+|\mathsf{Im}(\alpha_x)| \} / \sqrt{2}$.
The Hoeffding tail bound then yields
\begin{align}\label{ec}
\text{Pr} \left\{ \hat c > c + \delta_c \right\} \leq \exp{\left( - \frac{2 n \delta_c^2}{A^2 R^2} \right)}  
=: \epsilon_c \, .
\end{align}

Finally, consider the estimation of $P_0(R)$. This parameter is estimated by counting the number of times that a measurement output falls outside of the allowed range $\mathcal{R}(R)$. 
Bob can locally estimate this with the help of the auxiliary variables $S_i$, where $S_i=0$ if the $i$th signal falls inside the range, and $S_i=1$ otherwise.
Therefore, Bob's best estimate for $P_0(R)$ is
\begin{align}
    \hat P_0(R) = \frac{1}{n} \sum_{i=1}^n S_i \, .
\end{align}
This is the average of independent Bernoulli trials and therefore follows the Binomial distribution. A confidence interval can be obtained from the additive Chernoff bound:
\begin{align}
 & \text{Pr} \left\{ \hat P_0(R) < P_0(R) - \delta_P \right\} \nonumber \\
    & \leq \exp{\left[ - n D( P_0(R) - \delta_P \| P_0(R) )\right]} \, .
\end{align}
Applying the bound in Eq.~(\ref{KLbound}) we obtain
\begin{align}
    & \text{Pr} \left\{ \hat P_0(R) < P_0(R) - \delta_P \right\} \nonumber \\
    & \leq \exp{\left( - \frac{n \delta_P^2}{2P_0(R)(1-P_0(R)) } \right)} \\
    & \leq \exp{\left( - \frac{n \delta_P^2}{2P_0(R) }\right)} 
    =: \epsilon_P \, .
    \label{eP1}
\end{align}
We will require that the probabilities $\epsilon_v$, $\epsilon_c$, $\epsilon_P$ are much smaller than $1$, of the order of $10^{-10}$.

In summary, we have obtained that the following bounds,
\begin{align}
    \hat v & \geq v - \delta_v   \, , \label{condv} \\
    \hat c & \leq c + \delta_c   \, , \\
    \hat P_0(R) & \geq P_0(R) - \delta_P   \, , \label{condP}
\end{align}
hold true with almost unit probability (larger than $1-\epsilon_\text{PE}$, where $\epsilon_\text{PE} = \epsilon_v + \epsilon_c + \epsilon_P$ follows from an application of   the union bound).
For simplicity we put $\epsilon_v = \epsilon_c = \epsilon_P = \epsilon_\text{PE}/3$.
By inverting Eq.~(\ref{ec}), we obtain 
\begin{align} \label{delta_c}
    \delta_c & = AR \sqrt{\frac{\ln{(3/\epsilon_\text{PE})}}{2n}}  \, .
\end{align}
From Eqs.~(\ref{ev1}) and (\ref{eP1}) we obtain the following conditions for $\delta_v$ and $\delta_P$:
\begin{align}
\delta_v & = R \sqrt{ \frac{2 v \ln{(3/\epsilon_\text{PE})}}{n} } \, , \\
\delta_P & = \sqrt{ \frac{2 P_0(R) \ln{(3/\epsilon_\text{PE})}}{n} } \, .
\end{align}
To estimate these quantities we apply the inequalities (\ref{condv}), (\ref{condP}):
\begin{align}
\delta_v & \leq R \sqrt{ \frac{2 (\hat v +\delta_v) \ln{(3/\epsilon_\text{PE})}}{n} } \, , \\
\delta_P & \leq \sqrt{ \frac{2 (\hat P_0(R) +\delta_P) \ln{(3/\epsilon_\text{PE})}}{n} } \, .
\end{align}
Finally, solving for $\delta_v$ and $\delta_P$ we obtain
\begin{align}
\delta_v & \leq R \sqrt{ \frac{2 \hat v \ln{(3/\epsilon_\text{PE})} }{n} + \left( \frac{R \ln{(3/\epsilon_\text{PE})}}{n}\right)^2} \nonumber \\
& \phantom{=}~+  \frac{ R^2\ln{(3/\epsilon_\text{PE})}}{n} \, , 
\label{delta_v}\\
    \delta_P & \leq \sqrt{ \frac{2 \hat P_0(R) \ln{(3/\epsilon_\text{PE})} }{n} + \left(\frac{\ln{(3/\epsilon_\text{PE})}}{n}\right)^2} +  \frac{\ln{(3/\epsilon_\text{PE})}}{n} \, .
    \label{delta_P}
\end{align}

In conclusion, the non-asymptotic secret key rates are obtain using the formula in Eq.~(\ref{r_n}), where the parameter $\gamma_B(\tau)$ and $\gamma_{AB}(\tau)$ are obtained by solving the semi-definite programs (\ref{SDP-upperbound-finite}), (\ref{SDP-lowerbound-finite}) with the replacements
\begin{align}
    v & \to \hat v + \delta_v \, , \\
    c & \to \hat c - \delta_c \, , \\
    P_0(R) & \to \hat P_0(R) + \delta_P \, ,
\end{align}
and $\delta_v$, $\delta_c$, $\delta_P$ bounded as in Eqs.~(\ref{delta_c}), (\ref{delta_v}), (\ref{delta_P}).
The key rate obtained in this way is secure up to probability not larger than $\epsilon' = \epsilon_\text{s} + \epsilon_\text{h} + \epsilon_\text{PE}$.

\begin{figure}[t!]
\centering
\subfigure{\includegraphics[width=0.5\textwidth]{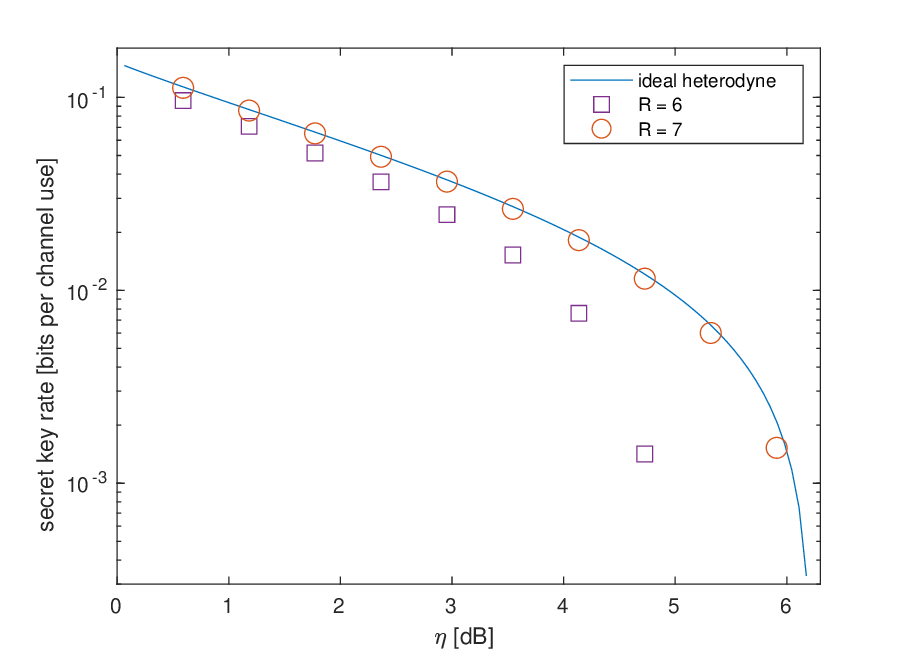}} \\
\subfigure{\includegraphics[width=0.5\textwidth]{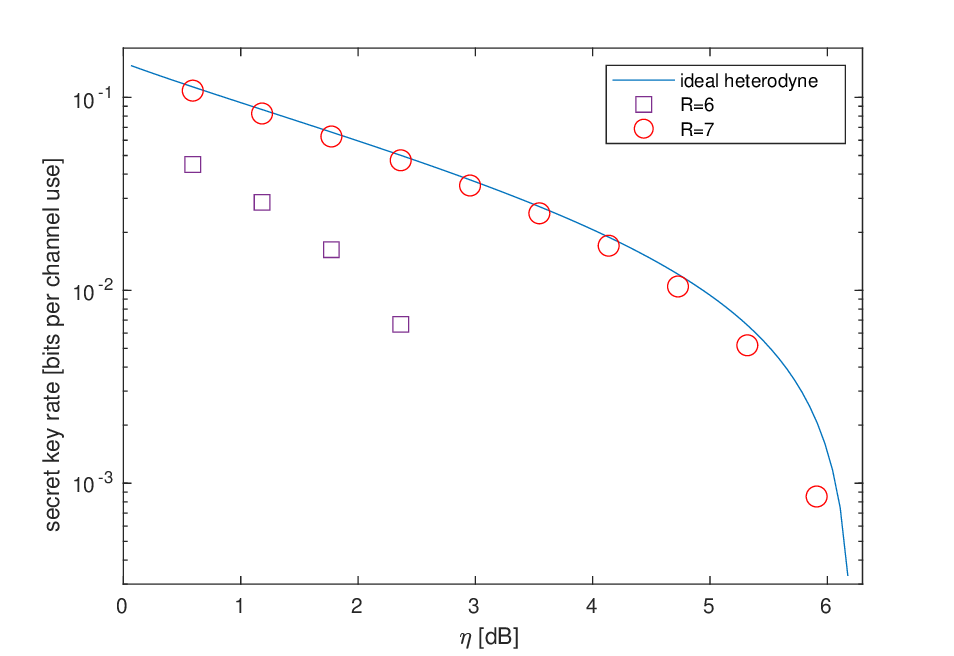}}
\caption{
Asymptotic secret key rates versus channel loss for QPSK encoding, for collective attacks in the limit of $n \to \infty$.
The channel parameters are $|\alpha| = 0.5$, $u = 0.001$, and $\xi = 0.97$.
The solid lines show the theoretical rate expected for ideal heterodyne detection, from Ref.~\cite{Lev21}.
For non-ideal heterodyne, the key rate is computed for $d=16$ and $R=6$ (squares) and $R=7$ (circles).
Top figure: the key rate is obtained by truncating and solving the \textit{infinite-dimensional} semi-definite programs (\ref{SDP-upperbound}) and (\ref{SDP-lowerbound}).
Bottom figure: the key rate is obtained by solving the \textit{finite-dimensional} semi-definite programs (\ref{SDP-upperbound-finite}) and (\ref{SDP-lowerbound-finite}).
}
\label{fig:channel_asym}
\end{figure}

\section{QPSK: secret key rates}\label{Sec:QPSK}
Our theoretical analysis applies to any DM protocol.
As a concrete example, we describe the application of our theory to QPSK encoding,
where $\alpha_x = \alpha i^x$ and $\mathcal{P}_x=1/4$, for $x=0,1,2,3$.
To align with the symmetry of our model of realistic heterodyne detection, we set $\alpha = |\alpha| e^{i \pi/4}$.

We have 
\begin{align}
    \alpha_x = |\alpha| e^{i \pi/4} i^x 
             = |\alpha| \frac{ \pm 1 \pm i}{\sqrt{2}} \, .
\end{align}
From this we obtain
\begin{align}
A = \frac{1}{\sqrt{2}} \max_x \left\{ |\mathsf{Re}(\alpha_x)| + |\mathsf{Im}(\alpha_x)| \right\} 
= |\alpha| \, ,
\end{align}
and
\begin{align}
    \| \mathcal{C} \|_\infty 
    \leq \frac{1}{2} \sup_{x,j,k} \left| \bar \alpha_x \beta_{jk} +  
    \alpha_x \bar \beta_{jk}\right| 
    \leq |\alpha| R \, .
\end{align}

For the sake of presentation, we assume a Gaussian channel from Alice to Bob, characterised by the loss factor $\eta \in [0,1]$ and the excess noise variance $u \geq 0$.
Given that $a$, $a^\dag$ are the canonical annihilation and creation operators on Alice's input mode, and $b$, $b^\dag$ on Bob's output mode, a Gaussian channel (in the Heisenberg picture) is a map of the form
\begin{align}
    b & \to \sqrt{\eta} \, a + \sqrt{1-\eta} \, e + w \, , \\
    b^\dag & \to \sqrt{\eta} \, a^\dag + \sqrt{1-\eta} \, e^\dag + \bar w \, , 
\end{align}
where $e$, $e^\dag$ are the canonical operators associated to an auxiliary vacuum mode, and $w$ is a Gaussian random variable with zero mean and variance $u$.
Assuming this form for the channel from Alice to Bob, we can explicitly compute the expected asymptotic values of the constraint parameters $v$, $c$, and $P_0(R)$, and then solve the semi-definite programs to estimate the CM elements $\gamma_B(\tau)$, $\gamma_{AB}(\tau)$. 
(More details are discussed in Appendix \ref{App:QPSK}.)
%

\begin{figure}[t!]
\includegraphics[width=1\linewidth]{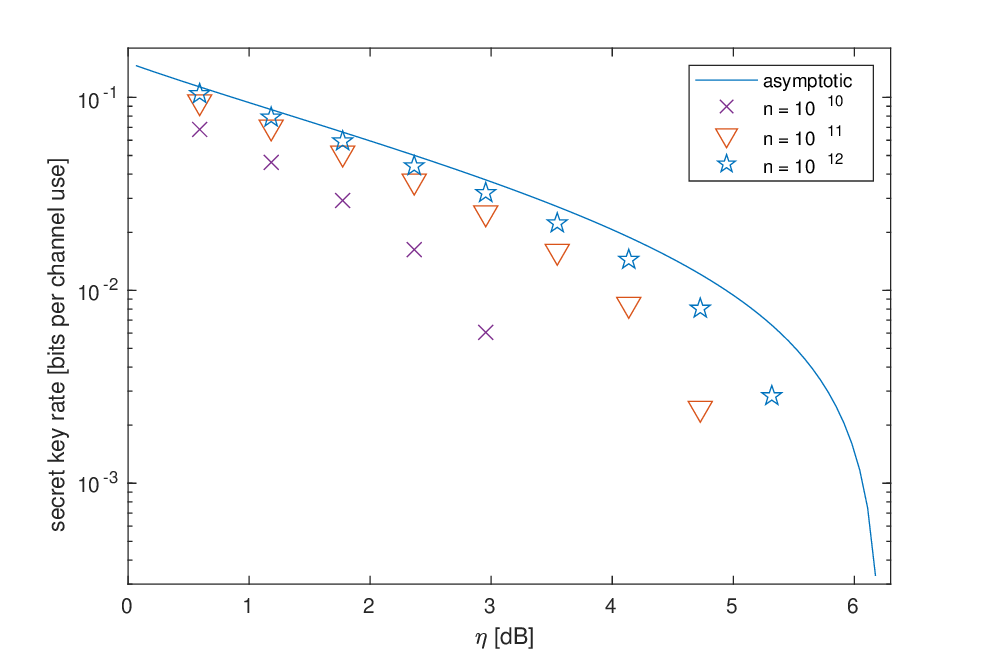}
\caption{
Composable secret key rates versus channel loss for QPSK encoding, for collective attacks in the regime of finite $n$.
The channel parameters are $|\alpha| = 0.5$, $u = 0.001$, $\xi = 0.97$.
The solid line is the theoretical rate expected in the asymptotic limit of $n \to \infty$ and for ideal heterodyne detection, from Ref.~\cite{Lev21}.
 For non-ideal heterodyne, the expected rate is computed from the \textit{finite-dimensional} semi-definite programs, and by taking into account finite-size corrections as described in Section \ref{Sec:finsize}.
 For $d=16$ and $R=7$, the plot shows the results for 
$n=10^{10}$ (crosses), 
$n=10^{11}$ (triangles), 
$n=10^{12}$ (stars).
The error parameters are $\epsilon_\text{h} = \epsilon_\text{s} = \epsilon_\text{PE} = 10^{-10}$.}
\label{fig:channel_finsize}
\end{figure}

The computed secret key rates (measured in bits per channel use, i.e., per mode) are shown in Figs.~\ref{fig:channel_asym}-\ref{fig:channel_finsize} versus the loss $\eta$, expressed in decibels. The other parameters of the protocol are fixed as $|\alpha| = 0.5$, and $u = 0.001$.

Figure~\ref{fig:channel_asym}(top) is obtained by solving the semi-definite programs (\ref{SDP-upperbound}) and (\ref{SDP-lowerbound}), which are defined in an infinite-dimensional Hilbert space.
To find a solution, we truncate the Hilbert space.
The figure shows that, as expected, by increasing $R$, and for $d$ large enough, the secret key rate converges towards the value expected for ideal heterodyne detection (which has been recently computed in Ref.~\cite{Lev21}). 
Our theory allows us to rigorously compute the deviation from this ideal rate.

Figure~\ref{fig:channel_asym}(bottom) is obtained by solving semi-definite programs (\ref{SDP-upperbound-finite}) and (\ref{SDP-lowerbound-finite}), which are defined in a finite-dimensional Hilbert space. 
In this case, a solution can be found without arbitrary truncation of the Hilbert space.
Compared with Fig.~\ref{fig:channel_asym}(top), we note that the secret key rate is reduced, especially if the value of $R$ is not large enough. 
This is due to the term proportional to $\| \mathcal{C}\|_\infty$ introduced in constraints of the semi-definite programs to account for the projections into the finite-dimensional space (therefore, an improved key rate can be obtained with a better bound for $\| \mathcal{C}\|_\infty$). 
However, already for $R=7$ the difference with the solution of the infinite-dimensional problem is relatively small.

Figure~\ref{fig:channel_finsize} is obtained by solving finite-dimensional semi-definite programs and including the finite-size corrections in the constraints, as discussed in Section \ref{Sec:finsize}.
For the sake of illustration, the calculations have been done by putting the best estimates of the parameters equal to the expected values, i.e.,
the semi-definite programs (\ref{SDP-upperbound-finite}), (\ref{SDP-lowerbound-finite}) are solved with the replacements
\begin{align}
    v & \to v + \delta_v \, , \\
    c & \to c - \delta_c \, , \\
    P_0(R) & \to P_0(R) + \delta_P \, .
\end{align}
The error parameters are $\epsilon_\text{h} = \epsilon_\text{s} = \epsilon_\text{PE} = 10^{-10}$.
The figure shows that a non-zero secret key rate is obtained when the block size is about $n=10^{10}$ or larger.
The dominant finite-size corrections are due to $\delta_v$ and $\delta_c$. This means that an improved key rate could be obtained by using tighter confidence intervals for the estimation of these parameters.
This, in turn, would allow us to reduce the block-size without compromising composable security.

\section{Conclusions} \label{Sec:end}

In CV QKD information is decoded by a coherent measurement of the quantum electromagnetic field, i.e., homodyne or heterodyne. 
These are mature technologies and represent the strategic advantage of CV QKD over discrete-variable architectures.
This applies to both continuous \cite{GG02,Weedbrook,Lev2015,Lev2017} and discrete modulation protocols \cite{Lev11,Lev19,Lut19,Lupo20,Wilde,Lut21,Lev21}.
Ideal homodyne and heterodyne detection, which are measurements of the quadratures of the field, possess a continuous symmetry that plays a central role in our theoretical understanding of CV QKD. 
%
However, this symmetry is broken in real homodyne and heterodyne dectection that are implemented in actual experiments \cite{Jouguet,mioQKD}.
%
While it is expected that, in practice, these measurements are well approximated by their idealised models in some regimes, a quantitative assessment of the error introduced by this approximation, and of its impact on the secret key rate, has so far been elusive.
Here we have filled this gap and presented a theory to quantify the security of CV QKD with real, imperfect, heterodyne detection. 
%
%
%
Within this theory we have established the composable security of DM CV QKD in the non-asymptotic regime. To the best of our knowledge this is the first result obtained in this direction, as previous works only considered asymptotic, non-composable security~\cite{Lut19,Lev19,Lut21,Lev21}.
Extension to most general attacks, which in principle can be obtained through a de Finetti reduction, remains an open problem.

In this paper, we have extended the approach of Ref.~\cite{Lev19}, in which one first estimates the covariance matrix of the quadratures, and then obtain a bound on the key rate using the property of extremality of Gaussian states. However, our theory can also be applied to the method of Refs.~\cite{Lut19,Lut21}, in which one uses the measured data to bound the key rate directly through non-linear semi-definite programming. 
%
We have focused on a particular kind of non-ideality in detection, but our approach can be applied to other non-idealities in both detection and in state preparation. Examples of these non-idealities include non-linearities in the analog-to-digital converter~\cite{mioQRNG} and noise in the state preparation~\cite{Lev21}. 
In principle, accounting for experimental imperfections in the security analysis mitigates the threat from side-channel attacks.
Our approach may also be extended to measurement-device-independent QKD~\cite{XXX,MDI1,MDI2}, which protects against unknown side-channel attacks on the detectors.
%
The results presented here are not only conceptually important, but will also enable secure, practical, and reliable DM CV QKD. 
In fact, to obtain reliable bounds on the secret key rates, the practitioner of CV QKD needs to carefully assess, in a composable way, finite-size effects as well as the impact of non-idealities in the measurement devices, including but not limited to, the effects of finite range and precision considered in this work.

\section*{Acknowledgments}

This work was supported by the EPSRC Quantum Communications Hub, Grant No.\ EP/T001011/1.
Y.O. is supported in part by NUS startup grants (R-263-000-E32-133 and R-263-000-E32-731), and the National Research Foundation, Prime Minister ’s Office, Singapore and the Ministry of Education, Singapore under the Research Centres of Excellence programme.

\appendix

\section{Holevo information}\label{App:Holevo}

Consider a two-mode state $\rho_{AB}$ shared between Alice and Bob.
We denote 
$a$, $a^\dag$, and 
$b$, $b^\dag$ 
the annihilation and creation operators on Alice's and Bob's mode, respectively.
Their local quadrature and phase operators are 
$q_A = (a+a^\dag)/\sqrt{2}$,
$p_A = (a-a^\dag)/(i\sqrt{2})$,
$q_B = (b+b^\dag)/\sqrt{2}$,
$p_B = (b-b^\dag)/(i\sqrt{2})$.
The symmetrically ordered CM $\gamma'(\rho)$ of the two-mode state $\rho$ is
defined as
\begin{align}
    \text{Tr} \left[ \rho \left(
    \begin{array}{cccc}
    q_A^2  & \Sigma(q_A,p_A) & q_A q_B & q_A p_B\\
    \Sigma(q_A,p_A) & p_A^2 & p_A q_B & p_A p_B \\
    q_B q_A & q_B p_A & q_B^2 & \Sigma(q_B,p_B) \\
    p_B q_A & p_B p_A & \Sigma(q_B,p_B) & p_B^2
    \end{array}
    \right)\right] \, ,
\end{align}
where $\Sigma(x,y) := (x y + y x )/2$.
The CM can be written in a block form as
\begin{align}
    \gamma'(\rho)= \left(
    \begin{array}{cc}
    A            & C \\
    C^\mathsf{T} & B 
    \end{array}
    \right) \, ,
\end{align}
where $A$, $B$, $C$ are $2 \times 2$ matrices.
We denote as $\nu_+$ and $\nu_-$ the symplectic eigenvalues of $\gamma'(\rho)$.
When Bob measures his mode by ideal heterodyne detection, the conditional state of Alice has CM
\begin{align}
    \gamma'(\rho_{A|B}) = 
    A - C (B+1/2)^{-1} C^\mathsf{T} 
    \, .
\end{align}
We denote $\nu_0$ as the symplectic eigenvalue of $\gamma(\rho_{A|B})$.

The property of extremality of Gaussian states yields the following bound on the Holevo information:
\begin{align}
    \chi(Y;E)_{\rho} \leq F_\chi(\gamma'(\rho)) \, ,
\end{align}
where
\begin{align}
    F_\chi(\gamma'(\rho)) = g(\nu_+-1/2) + g(\nu_--1/2) - g(\nu_0-1/2) \, ,
\end{align}
and for any $x>0$ the function $g$ is defined as
\begin{align}
    g( x ) := (x+1) \log_2{(x+1)} - x \log_2{x}
    \, ,
\end{align}
and $g(x):=0$ if $x = 0$.

It is possible to show \cite{Lev2015} that the function $F_\chi$ increases if we replace $\gamma'(\rho)$ with the matrix $\gamma(\rho)$
\begin{align}
    \text{Tr} \left[ \rho \left(
    \begin{array}{cccc}
    \Sigma(q_A^2,p_A^2) & 0 & \Delta & 0 \\
    0 & \Sigma(q_A^2,p_A^2) & 0 & -\Delta \\
    \Delta & 0 & \Sigma(q_B^2,p_B^2) & 0 \\
    0 & -\Delta & 0 & \Sigma(q_B^2,p_B^2)
    \end{array}
    \right)\right] \, ,
\end{align}
where $\Delta := (q_A q_B - p_A p_B) /2$.
From this we obtain the bound
\begin{align}
    \chi(Y;E)_{\rho} \leq F_\chi(\gamma(\rho)) \, ,
\end{align}

Note that
\begin{align}
\gamma_A(\rho) & := \frac{1}{2} \mathrm{Tr}[ ( a^\dag a + a a^\dag )  \rho ]
= \text{Tr} [\rho \Sigma(q_A^2,p_A^2) ] \, , \\
\gamma_B(\rho) & := \frac{1}{2} \mathrm{Tr}[ ( b^\dag b + b b^\dag ) \rho ] = \text{Tr} [\rho \Sigma(q_B^2,p_B^2) ] \, , \\
\gamma_{AB}(\rho) & := \frac{1}{2} \mathrm{Tr}[ ( a^\dag b^\dag + a b ) \rho ] 
= \text{Tr} [\rho \Delta ] \, .
\end{align}
Obviously, $F_\chi(\gamma(\rho))$ is a function of 
$\gamma_A(\rho)$,
$\gamma_B(\rho)$,
$\gamma_{AB}(\rho)$.
We therefore define 
\begin{align}
    f_\chi(\gamma_A(\rho),\gamma_B(\rho),\gamma_{AB}(\rho))  := F_\chi(\gamma(\rho)) \, .
\end{align}

\section{QPSK: EB representation}

In the PM representation, Alice prepares the state $|\alpha_x\rangle$ with probability $\mathcal{P}_x = 1/4$, for $\alpha_x = \alpha e^{i x \pi/2}$ and $x=0,1,2,3$, where we put $\alpha = |\alpha| e^{i \pi/4}$.

The average state prepared by Alice is
\begin{align}
\rho_{A'} = \frac{1}{4} \sum_x |\alpha_x \rangle \langle \alpha_x | \, .
\end{align}
We can expand this state in the number basis. Its $(n,n')$ entry is
\begin{align}
\rho_{A'}^{nn'} 
& = \frac{e^{-|\alpha|^2}}{4} \sum_x \frac{\alpha_x^n \bar\alpha_x^{n'}}{\sqrt{n! n'!}} \\
& = \frac{e^{-|\alpha|^2}}{4} \frac{\alpha^n \bar\alpha^{n'}}{\sqrt{n! n'!}}  \sum_x  e^{i (n-n') x \pi / 2 } \\
& = \frac{e^{-|\alpha|^2}}{4} \frac{\alpha^n \bar\alpha^{n'}}{\sqrt{n! n'!}}  
\left( 1 + e^{(n-n')\pi/2} \right)
\left( 1 + e^{(n-n')\pi}  \right)
\, .
\end{align}
That is, $\rho_{A'}^{nn'} = 0$ unless $n-n'$ is a multiple of $4$, in which case,
\begin{align}
\rho_{A'}^{nn'} 
= e^{-|\alpha|^2}
\frac{\alpha^n \bar\alpha^{n'}}{\sqrt{n! n'!}} \, .
\end{align}

As this state is invariant under rotation of $\pi/2$ in phase space, the eigenvectors have the form, for $y=0,1,2,3$,
\begin{align}
    |\phi_y\rangle = \sum_{n\ge 0} c_{y,n} |y + 4 n \rangle \, .
\end{align}

From 
\begin{align}
    \frac{1}{4} \sum_x |\alpha_x \rangle \langle \alpha_x | 
    = \sum_y \lambda_y |\phi_y\rangle \langle \phi_y | \, ,
\end{align}
we obtain
\begin{align}
\sqrt{\lambda_y} \, c_{y,n} = e^{-|\alpha|^2/2} \, \frac{\alpha^{y + 4n}}{\sqrt{(y+4n)!}}
\end{align}

By imposing normalisation, we find
\begin{align}
|\phi_y\rangle = \frac{e^{-|\alpha|^2/2} }{\sqrt{\lambda_y}} \sum_{n\ge 0} \frac{\alpha^{y + 4n}}{\sqrt{(y+4n)!}} 
| y + 4n\rangle \, ,
\end{align}
where
\begin{align}
\lambda_y = e^{-|\alpha|^2}  \sum_{n\ge 0} \frac{|\alpha|^{2(y + 4n)}}{(y+4n)!} \, .
\end{align}
Explicitly,
\begin{align}
    \lambda_0 & = \frac{e^{-|\alpha|^2}}{2} \left(  \cosh{\alpha^2} + \cos{\alpha^2}\right) \, , \\
    \lambda_1 & = \frac{e^{-|\alpha|^2}}{2} \left(  \sinh{\alpha^2} + \sin{\alpha^2}\right) \, , \\
    \lambda_2 & = \frac{e^{-|\alpha|^2}}{2} \left(  \cosh{\alpha^2} - \cos{\alpha^2}\right) \, , \\
    \lambda_3 & = \frac{e^{-|\alpha|^2}}{2} \left(  \sinh{\alpha^2} - \sin{\alpha^2}\right) \, .
\end{align}

We define the purification of the state $\rho_{A'}$ through its Schmidt decomposition,
\begin{align}
|\Psi\rangle_{AA'} = \sum_y \sqrt{\lambda_y} 
|\bar\phi_y\rangle |\phi_y\rangle \, ,
\end{align}
where
\begin{align}
|\bar \phi_y\rangle = \frac{e^{-|\alpha|^2/2} }{\sqrt{\lambda_y}} \sum_{n\ge 0} \frac{\bar\alpha^{y + 4n}}{\sqrt{(y+4n)!}} 
| y + 4n\rangle \, .
\end{align}

It is easy to check that
\begin{align}
|\alpha_x \rangle 
& = \sum_y e^{i x y \pi/2} \sqrt{\lambda_y} |\phi_y\rangle \, ,
\end{align}
which we can invert to obtain
\begin{align}
\sqrt{\lambda_y} |\phi_y\rangle = \sum_x \frac{e^{-i x y \pi/2}}{4} \, |\alpha_x \rangle \, .
\end{align}

We can then write
\begin{align}
|\Psi\rangle_{AA'} 
& = \sum_y \sqrt{\lambda_y} |\bar\phi_y\rangle |\phi_y\rangle \\
& = \sum_{xy} \frac{e^{-i x y \pi/2}}{4}  \, |\bar\phi_y\rangle |\alpha_x \rangle \\
& = \frac{1}{2} \sum_{x} |\psi_x\rangle |\alpha_x \rangle \, ,
\end{align}
where we have defined
\begin{align}
|\psi_x\rangle
= \frac{1}{2} \sum_{y} e^{-i x y \pi/2} |\bar\phi_y\rangle \, .
\end{align}

\section{Operators in the number representation}

We now express the operators that appear in our semidefinite programs in the basis $\smash{\{ |\psi_x\rangle \otimes |n\rangle\}_{x=0,\dots,3; n=0,\dots,\infty}}$, where $|n\rangle$'s are the number states of Bob's side, satisfying $b^\dag b |n\rangle = n |n\rangle$.

The operator $\rho_{B}$ is a density matrix of one bosonic mode. We can express it in the number basis, $\{ |n\rangle \}_{n=0,\dots,\infty}$, as
\begin{align}
    \rho_B = \sum_{nn'=0}^\infty \rho_{nn'} |n\rangle \langle n'| \, .
\end{align}
Similarly, $\rho_{AB}$ reads
\begin{align}
    \rho_{AB} = \sum_{xx'=0}^3 
    \sum_{nn'=0}^\infty \rho_{xx'nn'} |\psi_x\rangle \langle \psi_{x'}|
    \otimes 
    |n\rangle \langle n'| \, .
\end{align}

The operator $\Pi$ projects into the subspace with at most $N = \left \lfloor{2 R^2}\right \rfloor$ 
photons, i.e.,
\begin{align}
    \Pi = \sum_{n=0}^N |n\rangle \langle n| \, .
\end{align}
Therefore, 
\begin{align}
    \frac{1}{2} \Pi (b^\dag b + b b^\dag) \Pi = \sum_{n=1}^{N} \left( n + \frac{1}{2} \right) |n\rangle \langle n| \, .
\end{align}

The operator $\mathcal{V}$ is
\begin{align}
    \mathcal{V} = 
    \sum_{jk} |\beta_{jk}|^2 \int_{\mathcal{I}_{jk}} \frac{d^2\beta}{\pi} |\beta \rangle \langle \beta| 
    = \sum_{nn'=0}^\infty  \mathcal{V}_{nn'}
    |n\rangle \langle n'| \, ,
\end{align}
where
\begin{align}
    \mathcal{V}_{nn'} & = 
    \sum_{jk} |\beta_{jk}|^2 
    \int_{\mathcal{I}_{jk}} 
    \frac{d^2\beta}{\pi}  
    e^{-|\beta|^2} \frac{\beta^n \bar\beta^{n'}}{\sqrt{n! n'!}} \, .
\end{align}
Note that by symmetry, $\mathcal{V}_{nn'}=0$ unless $n-n'$ is multiple of $4$. Also by symmetry, $\mathcal{V}$ is a real matrix in the Fock basis.

Similarly, we have 
\begin{align}
    \mathcal{U} = 
    \int_{\beta \in \mathcal{R}(R)} 
    \frac{d^2\beta}{\pi} |\beta \rangle \langle \beta| 
    = \sum_{n,n'=0}^\infty  \mathcal{U}_{nn'}
    |n\rangle \langle n'| \, ,
\end{align}
with
\begin{align}
    \mathcal{U}_{nn'} & = 
    \int_{\beta \in \mathcal{R}(R)} 
    \frac{d^2\beta}{\pi}  
    e^{-|\beta|^2} \frac{\beta^{n} \bar\beta^{n'}}{\sqrt{n!n'!}} 
    \, .
\end{align}



The covariance operator in the objective function reads
\begin{align}
&\frac{1}{2} (a \Pi b \Pi + a^\dag \Pi b^\dag \Pi) \notag\\
 =& \frac{1}{2}\sum_{xx'=0}^3 \sum_{n=1}^{N} \sqrt{n} \langle \psi_x | a | \psi_{x'} \rangle |\psi_x\rangle \langle \psi_{x'}|\otimes |n-1\rangle \langle n|\notag\\
&\quad+ \sqrt{n}  \langle \psi_x | a^\dag | \psi_{x'} \rangle |\psi_{x}\rangle \langle \psi_{x'}| \otimes |n\rangle \langle n-1| \, , \\ 
=& \frac{1}{2}
\sum_{xx'=0}^3 \sum_{n=1}^{N} 
\sqrt{n}  \langle \psi_x | a | \psi_{x'} \rangle 
|\psi_x\rangle \langle \psi_{x'}|
\otimes |n-1\rangle \langle n| \notag\\
&\quad+ \sqrt{n}  
\langle \psi_{x'} | a | \psi_{x} \rangle^*
|\psi_{x}\rangle \langle \psi_{x'}|
\otimes |n\rangle \langle n-1| \, .
\end{align}

To compute this, first note that
\begin{align}
    \langle \bar \phi_{y-1} | a | \bar \phi_y\rangle 
    = 
    \bar\alpha \sqrt{\frac{\lambda_{y-1}}{\lambda_y}}  \, ,
\end{align}
from which we obtain
\begin{align}
    \langle \psi_{x} | a | \psi_{x'} \rangle 
    & = \frac{1}{4} \sum_{yy'=0}^3 
    e^{ i y x \pi/2} 
    e^{- i y' x' \pi/2} 
    \langle \bar \phi_y | a | \bar\phi_{y'}\rangle \\
    & = \frac{1}{4} \sum_{y=0}^3 
    e^{ i (y-1) x \pi/2} 
    e^{- i y x' \pi/2}
      \langle \bar\phi_{y-1} | a | \bar\phi_{y}\rangle \\
     & = \bar\alpha \, \frac{1}{4} \sum_{y=0}^3 
    e^{ i (y-1) x \pi/2} 
    e^{- i y x' \pi/2}
     \sqrt{\frac{\lambda_{y-1}}{\lambda_y}} \\
     & = \bar\alpha_x \, \frac{1}{4} \sum_{y=0}^3 
    e^{ i y (x-x') \pi/2}
     \sqrt{\frac{\lambda_{y-1}}{\lambda_y}}.
\end{align}

Finally, the operator $\mathcal{C}$ has components
\begin{align}
\mathcal{C}_{xx'nn'} 
     & = \frac{1}{2} \,
     \delta_{xx'} \bar\alpha_x 
    \hspace{-0.1cm}
    \sum_{j,k=1}^d
    \beta_{jk} 
    \int_{\mathcal{I}_{jk}}
    \hspace{-0.1cm}
    \frac{d^2 \beta}{\pi} 
    e^{-|\beta|^2} 
    \frac{\beta^n \bar \beta^{n'}}{\sqrt{n!n'!}} \nonumber \\
    & \phantom{=}~+
    \frac{1}{2} \, \delta_{xx'} \alpha_x 
    \sum_{j,k=1}^d
    \bar \beta_{jk}
    \int_{\mathcal{I}_{jk}}
    \hspace{-0.1cm}
    \frac{d^2 \beta}{\pi} 
    e^{-|\beta|^2} 
    \frac{\beta^n \bar \beta^{n'}}{\sqrt{n!n'!}} \, .
\end{align}

The operator can thus be written as
\begin{align}
     \mathcal{C}
     = \frac{1}{2} \left( \mathcal{A}^\dag \otimes \mathcal{B} + \mathcal{A} \otimes \mathcal{B}^\dag \right) \, ,
\end{align}
where
\begin{align}
    \mathcal{A} & = \sum_{x=0}^3 \alpha_x |\psi_x\rangle \langle \psi_x| \, , \\
    \mathcal{B} & = \sum_{nn'}    \sum_{j,k=1}^d
    \beta_{jk} 
    \int_{\mathcal{I}_{jk}}
    \hspace{-0.1cm}
    \frac{d^2 \beta}{\pi} 
    e^{-|\beta|^2} 
    \frac{\beta^n \bar \beta^{n'}}{\sqrt{n!n'!}} \,
    |n\rangle \langle n'| \, .
\end{align}

Note that, by symmetry, 
$[\mathcal{B}]_{nn'}=0$ for $n-n'$ even.
Also by symmetry, the entries of $\mathcal{C}$ are all real.

\section{Semidefinite programming}\label{App:SDP}

In the main body of the paper we have formulated the following optimisation problems
\begin{maxi}
{\rho \ge 0}{ \frac{1}{2} \< \Pi (b^\dag b + b b^\dag )\Pi , \rho \>} {}{}
\addConstraint{ \< \mathcal V, \rho\>  }{\leq v}{}
\addConstraint{ \< \mathcal{U}, \rho\> }{\geq 1-P_0(R)}{}
\addConstraint{ \< I, \rho\>  }{= 1}{}. \label{SDP-upperbound-app}
\end{maxi} 
and
\begin{mini}
{\rho \ge 0}{ \frac{1}{2} \<a \Pi b \Pi + a^\dag \Pi b^\dag \Pi, \rho \> }{}{}
\addConstraint{ \< I\otimes\mathcal V, \rho\>  }{\leq v}{}
\addConstraint{ \< \mathcal C, \rho\>  }{\geq c}{}
\addConstraint{ \< I\otimes\mathcal{U}, \rho\>  }{\geq 1-P_0(R)}{}
\addConstraint{ \mathrm{Tr}_B(\rho)  }{= \sigma}{}
\addConstraint{ \<I , \rho\>  }{= 1}{}, \label{SDP-lowerbound-app}
\end{mini} 
where $\<A, X\> =  {\rm Tr}(A^\dagger X)$ denotes the Hilbert Schmidt inner product.
To derive the corresponding dual programs which will be more numerically efficient to evaluate, we revisit duality theory for SDP with mixed constraints. 
Given any semidefinite program of the form
\begin{mini}
{X \ge 0}{\< C, X \>}{}{}
\addConstraint{ \<A_i, X\> }{ \le a_i}{}
\addConstraint{ \<B_j, X\> }{ = b_j}{} ,
\end{mini} 
where $C,A_i$ and $B_j$ are Hermitian matrices,
the Lagrangian is given by
\begin{align}
    L = \<C,X\> +\sum_{i} y_i \left( \<A_i , X\> - a_i \right)
    + \sum_{j} z_j \left( \<B_j , X\> - b_j \right),
\end{align}
where $y_i \ge 0, z_i \in \mathbb R$.
By linearity of inner products, we can rewrite the Lagrangian as
\begin{align}
    L = \<C + \sum_i y_i A_i + \sum_j z_j B_j,X\> - \sum_{i} y_i a_i - \sum_{j}z_j b_j.
\end{align}
The Lagrange dual is then given by
\begin{maxi}
{y_i \ge 0, z_j \in \mathbb R}
{ - \sum_{i} y_i a_i - \sum_{j}z_j b_j }{}{}
\addConstraint{ C + \sum_i y_i A_i+ \sum_j z_j B_j    }{\ge 0}{}.
\end{maxi} 

The Lagrange dual of (\ref{SDP-upperbound-app}) is thus given by 
\begin{mini}
{y_1, y_2 \ge 0, z \in \mathbb R}
{ y_1 v - y_2 (1-P_0(R)) + z \phantom{==========}}{}{}
\addConstraint{ - \frac{1}{2}\Pi (b^\dag b + b b^\dag ) \Pi + y_1 \mathcal V - y_2 \mathcal{U} + z I    }{\ge 0}{}. \label{SDP-upperbound-finite_relaxation_dual}
\end{mini} 
Strong duality in this case holds because the inequality constraints can be strictly feasible, and the Slater constraint qualification holds.

The Lagrange dual of \eqref{SDP-lowerbound-app} can be written as
\begin{maxi}
{\substack{y_1,y_2,y_3 \ge 0,y_4 \in \mathbb R\\
z_{h,k} \in \mathbb R}}
{  y_1 c - y_2 v + y_3 (1-P_0(R)) - y_4 - \phi(z) \phantom{====}}{}{}
\addConstraint{ 
\frac{1}{2} ( a \Pi b \Pi + a^\dagger \Pi b^\dagger \Pi ) + \kappa(y,z)
}{\ge 0}{}
.
\label{SDP-lowerbound-finite_relaxation_dual}
\end{maxi} 
where
\begin{align}
\kappa(y,z) & = - y_1\mathcal C + y_2\mathcal V - y_3\mathcal{U} + y_4 I + \sum_{h, k} z_{h,k} Z_{h,k} \, , \\
\phi(z) & = \sum_{h \ge k} z_{h,k} {\rm Re}(\sigma_{h,k} )
+ \sum_{h<k} z_{h,k} {\rm Im}(\sigma_{h,k}) \, ,
\end{align}
and
$Z_{h,k} = E_{h,k} \otimes I_B  $ when $h \ge k$ and
$ Z_{h,k} = F_{k,h} \otimes I_B  $ when $h<k$, with
\begin{align}
    E_{h,k} & = 
    \frac{ |k\>\<h| + |h\>\<k| }{2} \, , \\
    F_{h,k} & = 
    i \, \frac{ |k\>\<h| - |h\>\<k| }{2} \, .
\end{align}

\begin{figure}[t!]
\includegraphics[width=0.99\linewidth]{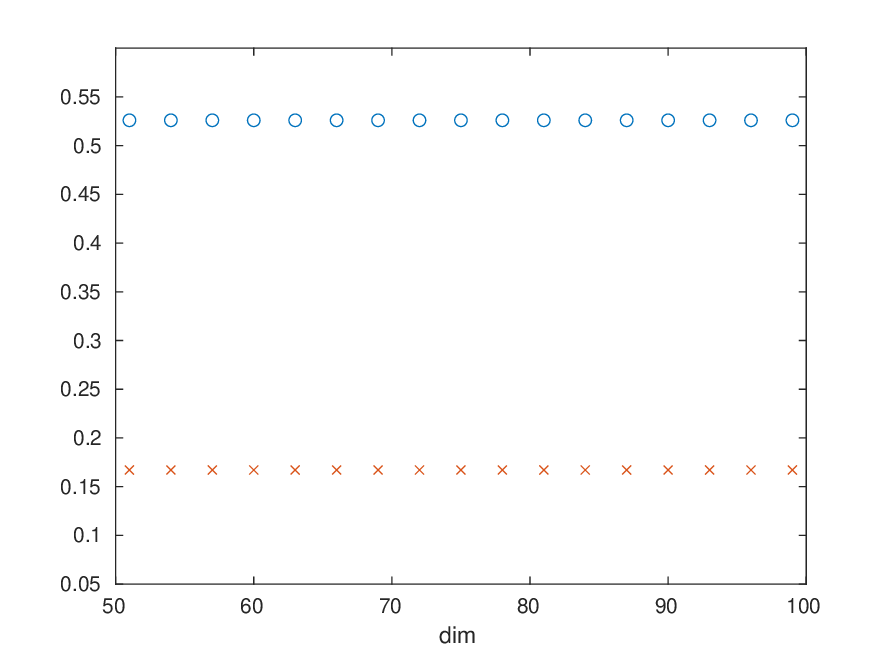}
\caption{
Optimal values plotted versus the Hilbert space cutoff $\text{dim}$, 
for (\ref{SDP-upperbound-finite_relaxation_dual}) (circles) 
and (\ref{SDP-lowerbound-finite_relaxation_dual}) (crosses),
and $R=7$, $d=16$, $\alpha=0.5$, $\eta = 0.1$, $u=0.001$.
The optimal values are largely independent of $\text{dim}$ as long as $\text{dim} \geq \lfloor R^2 \rfloor + 1$.}
\label{fig:R7}
\end{figure}

To solve numerically these optimisation problems we need to impose a cutoff to Bob's Hilbert space, and work within a finite dimensional space of dimensions $\text{dim}$, containing no more than $(\text{dim}-1)$ photons on Bob's side.
The value of $\text{dim}$ can be arbitrarily large, as long as it is larger than $N + 1$, where $N = \lfloor 2 R^2 \rfloor$ is determined by the rank of the projector $\Pi$. However, our numerical results suggest that it is sufficient to put $\text{dim} = N + 1$. 
As an example, Fig.~\ref{fig:R7} shows the optimal values for QPSK encoding, and for the optimisation problems (\ref{SDP-upperbound-finite_relaxation_dual}) 
and (\ref{SDP-lowerbound-finite_relaxation_dual}), as a function of $\text{dim}$.

\section{QPSK: secret key rates} \label{App:QPSK}

As a concrete example, we apply our theory to QPSK encoding,
where $\alpha_x = \alpha i^x$ and $\mathcal{P}_x=1/4$, for $x=0,1,2,3$.
To align with the symmetry of our model of realistic heterodyne detection, we set
$\alpha = |\alpha| e^{i \pi/4}$.
We simulate a Gaussian channel from Alice to Bob, characterised by the loss factor $\eta \in [0,1]$ and the excess noise variance $u \geq 0$.

First, we compute the expected value for the mutual information,
\begin{align}
I(X;Y) = H(Y) - H(Y|X) \, ,
\end{align}
where $H(Y)$ is the entropy of Bob's measurement outcome, and $H(Y|X)$ is the conditional entropy for given input state prepared by Alice.
If Alice prepares the coherent state $|\alpha_x\rangle$, with $\alpha_x = ( q_x + i p_x)/\sqrt{2}$, then the state $\rho_B(x)$ received by Bob is described by the Wigner function $W_x(q,p)$, where
\begin{align}
    W_x(q,p) = \frac{1}{\pi(2u+1)} \, e^{ - \frac{ (q - \sqrt{\eta} \, q_x)^2 + (p - \sqrt{\eta} \, p_x)^2 }{2 (u+1/2)} } \, .
\end{align}
From this, we obtain the probability density of measuring $\beta = (q + i p)/\sqrt{2}$ by ideal heterodyne detection, 
\begin{align}
    \frac{1}{\pi} \langle \beta | \rho_B(x) | \beta \rangle 
    = \frac{1}{2\pi(u+1)} \, e^{ - \frac{ (q - \sqrt{\eta} \, q_x)^2 + (p - \sqrt{\eta} \, p_x)^2 }{2 (u+1)} } \, ,
\end{align}
and, in turn, the probability of measuring $\beta \in \mathcal{I}_{jk}$,
\begin{align}
    P_{jk|x} = \frac{1}{\pi} \int_{\beta \in \mathcal{I}_{jk}} 
    d^2 \beta
    \langle \beta | \rho_B(x) | \beta \rangle = P_{j|x} \, P_{k|x} \, ,
\end{align}
where
\begin{align}
    P_{j|x} & = \frac{1}{2} \text{erf}
    \left[ 
    \frac{(2+d-2j)R + d \sqrt{\eta} \, q_x}{d \sqrt{2(u+1)}}
    \right] \nonumber \\
    &\phantom{=}~ - \frac{1}{2} \text{erf}
    \left[ 
    \frac{(d-2j)R + d \sqrt{\eta} \, q_x}{d \sqrt{2(u+1)}}
    \right] \, .
\end{align}

For QPSK encoding, the conditional mutual information then reads ($\log$ in base $2$)
\begin{align}
    H(Y|X) = - \frac{1}{4} \sum_{x=0}^3 \sum_{j,k=1}^d P_{jk|x} \log{P_{jk|x}} \, .
\end{align}
The probability distribution of $Y$ is obtained by averaging over $X$,
$P_{jk} = \frac{1}{4} \sum_{x=0}^3 P_{jk|x}$,
and the entropy of $Y$ is
\begin{align}
    H(Y) = - \sum_{j,k=1}^d P_{jk} \log{P_{jk}}  \, .
\end{align}


Similarly, we compute the expected values for the estimated parameters $v$ and $c$.
We obtain
\begin{align}
    \mathrm{Tr}(\mathcal{C} \rho_{AB}) 
    & = \frac{1}{4}
    \sum_{x=0}^3 \sum_{j,k=1}^d 
    \frac{q_x q_j + p_x p_k}{2} 
    \,
     P_{jk|x} \\
\mathrm{Tr}(\mathcal{V} \rho_B ) 
    & = \sum_{j,k=1}^d \frac{q_j^2 + p_k^2}{2} \, P_{jk} \, .
\end{align}


\end{document}